\documentstyle[prl,aps,twocolumn,epsfig]{revtex}

\begin{document}
\twocolumn[\hsize\textwidth\columnwidth\hsize\csname @twocolumnfalse\endcsname

\title{Universal Phase Diagram for High-Piezoelectric Perovskite Systems}
\author{D. E. Cox, B. Noheda\thanks{%
To whom correspondence should be addressed
(e-mail: noheda@bnl.gov).}, and G. Shirane}
\address{Brookhaven National Laboratory, Upton, New York 11973,USA.}
\author{Y. Uesu, K. Fujishiro, and Y. Yamada.}
\address{Waseda University, Shinjuku-ku, Tokyo 169, Japan.}
\maketitle

\begin{abstract}
Strong piezoelectricity in the perovskite-type PbZr$_{1-x}$Ti$_{x}$O$_{3}$
(PZT) and Pb(Zn$_{1/3}$Nb$_{2/3}$)O$_{3}$-PbTiO$_{3}$ (PZN-PT) systems is
generally associated with the existence of a morphotropic phase boundary
(MPB) separating regions with rhombohedral and tetragonal symmetry. An x-ray
study of PZN-9\%PT has revealed the presence of a new orthorhombic phase at
the MPB, and a near-vertical boundary between the rhombohedral and
orthorhombic phases, similar to that found for PZT between the rhombohedral
and monoclinic phases. We discuss the results in the light of a recent
theoretical paper by Vanderbilt and Cohen, which attributes these
low-symmetry phases to the high anharmonicity in these oxide systems.

\end{abstract}
\vskip1pc]

\narrowtext

Ferroelectric perovskite-type materials are of great fundamental and
technological importance. Among such materials, the PZT and PZN-PT systems
exhibit unusually large piezoelectric coefficients; in particular,
extraordinarily high values have recently been reported for PZN-PT by Park
and Shrout when an electric field is applied along the pseudocubic [001]
direction \cite{Par1}, with $d_{33}> 2500$pC/N and strain values up to 1.7 $%
\%.$ These properties are an order-of-magnitude superior to those of PZT
ceramics, currently the materials of choice for a wide variety of
high-performance electromechanical devices, and make PZN-PT a promising
candidate for a new generation of such devices \cite{Ser}. In both PZT and
PZN-PT, the high piezoelectric coefficients have been associated with an MPB
which separates regions having tetragonal (T) and rhombohedral (R) symmetry
. However, the recent discovery of a new monoclinic phase in PZT at the MPB
has changed this picture dramatically \cite{Noh1,Noh2,Guo,Bel,Noh3}.The
phase diagrams for the two systems are shown in Fig. 1 \cite{Jaf,Kuw1}, with
the respective MPB's represented as heavy lines and the recently-discovered
monoclinic structure in PZT \cite{Noh1} as M$_A$. A key feature of this new
structure is that the polarization vector is no longer constrained to lie
along a symmetry axis, as in the rhombohedral (R) and tetragonal (T)
structures, but instead can rotate within the monoclinic plane \cite
{Noh2,Guo,Bel}.

In the present paper, we report the results of a high-resolution synchrotron
x-ray powder diffraction study of a sample doped with 9$\%$PT. This study
complements single-crystal investigations of 4.5\% and 8\% PT samples
carried out in collaboration with the Penn State group which will be
reported elsewhere. In particular, measurements made on rhombohedral 8\%PT
samples under an electric field have shown that an orthorhombic phase is
irreversibly induced by the field, indicative of the close proximity of a
rhombohedral-orthorhombic boundary in the phase diagram\cite{Noh4}. The
presence of some lower-symmetry phase at room temperature was also revealed
in a previous optical study of 9\%PT\cite{Fuj}, and the present x-ray
results show conclusively the presence of this new phase in the PZN-PT phase
diagram, which, in contrast to the case of PZT, has orthorhombic rather than
monoclinic symmetry.

\begin{figure}[tbp]
\epsfig{width=0.85 \linewidth,figure=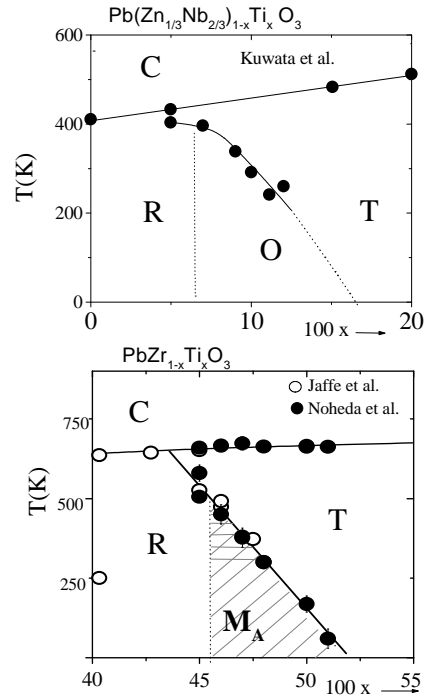}
\caption{ Phase diagrams for PZT (bottom) and PZN-PT (top) in the vicinity
of their respective MPB's, which are shown by the heavy lines\protect\cite
{Jaf,Kuw1}. C, R and T represent cubic, rhombohedral and tetragonal regions.
The diagonally-shaded M$_A$ area in the PZT diagram represents the stability
region of the recently-discovered monoclinic phase\protect\cite{Noh1}. The
cross-shaded area at the upper tip of the M$_A$ region represents a complex
area of phase coexistence\protect\cite{Noh3}. The proposed stability region
in the PZN-PT phase diagram for the new orthorhombic phase is labeled O.
Note that in both phase diagrams, the boundaries between the lower-symmetry
and R phases are nearly vertical.}

\end{figure}
\begin{figure}[tbp]
\epsfig{width=0.9 \linewidth,figure=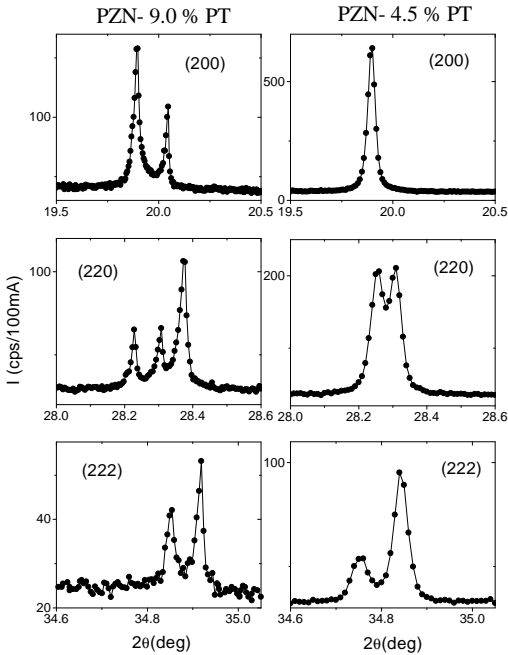}
\caption{High-resolution synchrotron x-ray diffraction data from
polycrystalline samples of PZN containing 9\% PT (left) and 4.5\% PT
(right), prepared as described in the text.}
\end{figure}

Compared to the PZT system PZN-PT has the great advantage that single
crystals are relatively easy to grow. A crystal of Pb(Zn$_{1/3}$Nb$_{2/3}$)O$%
_{3}$ containing 9 $\%$ PbTiO$_{3}$, identical to those described in ref. 
\cite{Mul}, was grown from a PbO flux. It is important to emphasize that in
order to obtain sharp powder diffraction peaks from a relaxor-type
ferroelectric is necessary to induce the ordered ferroelectric state by
application of an electric field to the as-grown crystal, and then to give
particular attention to the preparation of a suitable polycrystalline
sample, since the diffraction peak profiles can be seriously degraded due to
the effects of particle-size or microstrain broadening caused by excessive
grinding. This was accomplished by use of the following procedure. A small
fragment was chopped out of the central region of the crystal and lightly
crushed and ground in an agate mortar under acetone. The fraction of
crystallites retained between 325-400 mesh sieves (~$\sim $38-44 microns)
was loaded into a thin-walled glass capillary tube 0.2 mm in diameter which
was then sealed. The use of a very narrow-diameter sample is mandated by the
extremely high absorption coefficients of these materials in the accessible
range of wavelengths. High resolution powder data were collected at the
beamline X7A at the Brookhaven National Synchrotron Light Source with a
double-crystal Ge(111) monochromator set for a wavelength of ~$\sim $0.7 
\AA\ and a flat Ge(220) crystal-analyzer mounted in the path of the
diffracted beam. The instrumental resolution (full-width at half-maximum) in
this range is 0.005-0.01$^{o}$, an order-of-magnitude better than that of a
typical laboratory diffractometer. During data collection, the samples were
either rotated at about 1 Hz or rocked over several degrees, which is
essential to achieve powder averaging over crystallites of this size. The
peaks were found to be extremely sharp and well-resolved, as illustrated by
the data shown on the left side of Fig. 2. From the splitting and relative
intensities of the peaks it is easy to deduce that the sample has neither
rhombohedral nor tetragonal symmetry, nor does it consist of a mixture of
two such phases in coexistence. However, all the peaks can be indexed very
satisfactorily on the basis of a B-centered orthorhombic cell with $%
a_{o}=5.737$ $\AA $, $b_{o}=4.033$ $\AA $, $c_{o}=5.759$ $\AA $, similar to
that reported for BaTiO$_{3}$ in the temperature range 180-270 K\cite{Jon}.
For comparison, similar data obtained from a sample of PZN-4.5$\%$PT are
displayed on the right side of Fig. 2. These peaks are somewhat broader but
otherwise well-resolved; however in this case the symmetry is unambiguously
rhombohedral as previously reported for this composition\cite{Kuw1}, with $%
a=4.057$ $\AA $, $\alpha =89.89^{\circ }$. The results of the
temperature-dependence studies are presented in Fig.3. Between 350-450 K,
the symmetry was found to be tetragonal with the $c_{t}/a_{t}$ ratio
decreasing from 1.009 to 1.003, in agreement with previously reported results
\cite{Kuw1}. Below room temperature, the system remains orthorhombic down to
20 K, with a steady decrease in $b_{o}$ and a small increase in the
orthorhombic distortion $c_{o}/a_{o}$ from 1.0027 to 1.0041. In contrast to
BaTiO$_{3}$, there is no further transition to a rhombohedral phase.
Although the temperature increments are relatively coarse, the
orthorhombic-tetragonal phase transition is clearly first-order, as
reflected by a sharp discontinuity in the behavior of the $a_{o}$ and $c_{o}$
lattice parameters. Interestingly, though, there is no such discontinuity
between $b_{o}$ and $a_{t}$, which are perpendicular to the respective polar
axes. This behavior is consistent with a first-order transition in which the
polar axis jumps from pseudocubic [101] to [001].

In a previous investigation, the diffraction profiles from unpoled single
crystals of PZN-9$\%$PT were interpreted in terms of the coexistence of R
and T phases\cite{Ues}. However a recent analysis of the old data is in
complete agreement with the orthorhombic phase. As noted above, this
B-centered orthorhombic unit cell is similar to that induced by an electric
field in rhombohedral 8$\%$ PT\cite{Noh4}. As discussed in the latter paper,
it can also be described in terms of a primitive monoclinic cell, space
group Pm, with $a=c=4.062$ $\AA ,$ $b=4.033$ $\AA ,$ $\beta =90.15^\circ $.
Thus, although we find no evidence of a truly monoclinic distortion (since $%
a=c$ within the experimental error limits of $\sim $0.002 \AA ), the
orthorhombic phase can be regarded as a Pm monoclinic phase (M$_C$ in the
notation of ref. \cite{Van}) in the limit of $a=c$. However, this is not the
same type of monoclinic distortion (M$_A$ in this notation) found in the PZT
system\cite{Noh1,Noh2}, which has Cm symmetry.

\begin{figure}[tbp]
\epsfig{width=0.8 \linewidth,figure=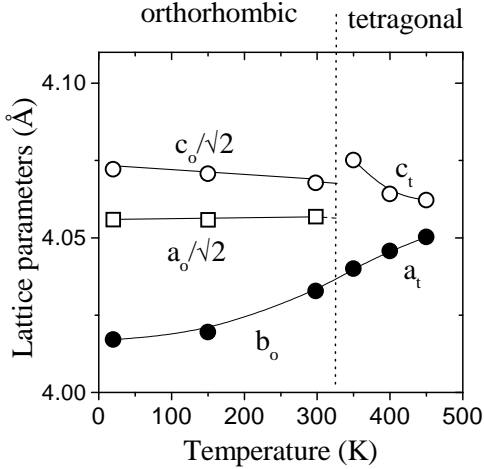}
\caption{ Fig.3. Variation of orthorhombic lattice parameters for PZN-9\%PT
as a function of temperature between 20-450 K. The sharp discontinuities in $%
a_{o}$ and $c_{o}$ with respect to $c_{t}$ between the orthorhombic and
tetragonal regions and the smooth variation of $b_{o}$ and at are consistent
with a jump of the polarization vector from orthorhombic $\left[ 001\right] $
to tetragonal $\left[ 001\right] $, i.e. from pseudocubic $\left[ 101\right] 
$ to $\left[ 001\right] $.}
\end{figure}

It is of great interest to view the above results in the light of the recent
paper by Vanderbilt and Cohen\cite{Van}, hereafter referred to as VC, in
which an extension of the Devonshire theory\cite{Dev} to eighth-order is
found to provide a natural description of the recently-discovered monoclinic
phase in the PZT system\cite{Noh1,Noh2}. The new theory also predicts the
existence of stability regions for two other types of monoclinic phase and
the nature of the boundaries between the various phases. In Fig. 4 we
reproduce a slightly modified version of the VC phase diagram in the space
of two dimensionless parameters, $\alpha $ and $\beta $, which contains
stability regions for three kinds of monoclinic phase, labelled M$_A$, M$_B$
and M$_C$, in addition to the T, R and O phases found in the sixth-order
theory. The monoclinic phase in the PZT system is of M$_A$ type (space group
Cm). As noted by VC, although $\alpha $ and $\beta $ cannot be related to
temperature or composition in any simple way, it is clear from the data
shown in the lower part of Fig. 1 that the MPB in PZT must lie close to the
triple-point connecting the R, M$_A$ and T phases. The PZN-PT system
evidently behaves in a quite different fashion. In this case, the discovery
of the new phase at 9$\%$ PT shows that there must be a narrow orthorhombic
region in the vicinity of the MPB, as shown schematically in the upper part
of Fig. 1. Then it is very plausible that the MPB must lie close to the
triple-point on the right-hand side connecting the R, O and T phases, and
that 9$\%$ PT must fall to the left of this triple-point. As can be seen in
Fig. 1, the only difference between the two phase diagrams is the structure
of the intermediate phase between R and T. As discussed by VC, the clear
signature from monoclinic M$_A$ places the PZT system on the phase diagram
in the region shown by the shaded area on the left side of Fig. 4. However,
the observed orthorhombic phase in PZN-PT does not uniquely locate the
system on this diagram, and our present conjecture is shown by the shaded
region on the right side of Fig. 4. We emphasize that so far we have only
studied the one composition with 9$\%$PT, and it is possible that
compositions between 10-12$\%$PT might show the monoclinic M$_C$ structure,
which would establish orthorhombic 9$\%$PT as the end member of a new
monoclinic system.
\begin{figure}[tbp]
\epsfig{width=0.7 \linewidth,figure=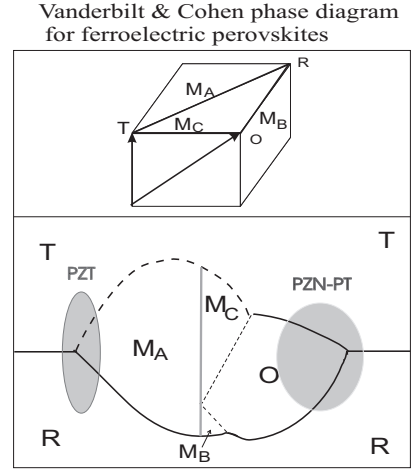}
\caption{Phase diagram for ferroelectric perovskites in the space of the
dimensionless parameters $\protect\alpha $ (vertical axis) and $\protect%
\beta $ (horizontal axis), which reflect the relative importance of the
coefficients of the fourth, sixth and eighth-order terms in the free-energy
expansion (after Vanderbilt and Cohen\protect\cite{Van}). In the top part 
of the figure, the T, O and R phases are
depicted as points at the corners of a cube for which the polarization
vector (shown by the heavy arrows for T and O) is constrained to lie along a
symmetry axis, and the M$_A$, M$_B$ and M$_C$ phases as symmetry lines for
which the polarization vector lies within a symmetry plane. }
\end{figure}

So far we have not addressed the important issue of the polarization path
under an applied electric field, a topic treated theoretically by Fu and
Cohen\cite{Fu} by applying first-principles calculations to BaTiO$_{3}$. An
experiment of this type carried out on PZN-9$\%$PT may shed new light on the
question as to whether the new orthorhombic phase in this material is
fundamentally different from the old orthorhombic phase in BaTiO$_{3}$. We
also anticipate that further insight will be obtained from theoretical
calculations of the stability of the E=0 phase, which were so useful in the
case of PZT\cite{Bel}. In conclusion, we propose that the appearance of a
lower-symmetry phase in the vicinity of a MPB over a narrow range of
composition and temperature, together with a near-vertical boundary between
the lower-symmetry and rhombohedral phases, are universal features of
highly-piezoelectric perovskite systems. This has important implications for
further improving the performance of electromechanical devices since it is
just at the MPB that the piezoelectric response is greatly enhanced in the
PZT and PZN-PT systems. We hope that this will stimulate further work to
synthesize new systems with MPB's and explore other regions of the phase
diagram in Fig. 4.

 \acknowledgments
The authors thank L.E. Cross, S-E. Park (Penn. State U.), D. Vanderbilt
(Rutgers U.) and Z-G. Ye (Simon Fraser U.) for many helpful discussions and
suggestions, and Y. Yamashita (Toshiba Research Center) for providing the
PZN-PT single crystals. Financial support from the U.S. Department of Energy
and the Japanese Ministry of Education is also gratefully acknowledged.

\end{document}